# Non-volatile heterogeneous III-V/Si photonics via optical charge-trap memory

Stanley Cheung*, Di Liang*, Yuan Yuan*, Yiwei Peng, Yingtao Hu, Geza Kurczveil, and Raymond G. Beausoleil

Hewlett Packard Enterprise, Large-Scale Integrated Photonics Lab, Milpitas, CA. 95035, USA
*stanley.cheung@hpe.com, di.liang@ieee.org, yuan.yuan@hpe.com

Abstract

We demonstrate, for the first time, non-volatile charge-trap flash memory (CTM) co-located with heterogeneous III-V/Si photonics. The wafer-bonded III-V/Si CTM cell facilitates non-volatile optical functionality for a variety of devices such as Mach-Zehnder Interferometers (MZIs), asymmetric MZI lattice filters, and ring resonator filters. The MZI CTM exhibits full write/erase operation (100 cycles with 500 states) with wavelength shifts of $\Delta\lambda_{non\text{-}volatile} = 1.16$ nm ($\Delta n_{eff,non\text{-}volatile} \sim 2.5 \times 10^{-4}$) and a dynamic power consumption < 20 pW (limited by measurement). Multi-bit write operation (2 bits) is also demonstrated and verified over a time duration of 24 hours and most likely beyond. The cascaded 2$^{nd}$ order ring resonator CTM filter exhibited an improved ER of ~ 7.11 dB compared to the MZI and wavelength shifts of $\Delta\lambda_{non\text{-}volatile} = 0.041$ nm ($\Delta n_{eff,\,non\text{-}volatile} = 1.5 \times 10^{-4}$) with similar pW-level dynamic power consumption as the MZI CTM. The ability to co-locate photonic computing elements and non-volatile memory provides an attractive path towards eliminating the von-Neumann bottleneck

## Introduction

Increasing workloads in data centers and high-performance computing are dominated by data and graph analytic applications aided by artificial intelligence (AI) [1,2]. Nowadays, these data-intensive applications are handled by graphical processing unit (GPU)-based artificial neural networks (ANNs) using traditional von Neumann computing architecture [2]. The majority of energy consumption in an electrical ANN comes from matrix vector multiplication (MVM) and data movement operations. The well-known von Neumann bottleneck also fundamentally limits the speed and energy-efficiency of the data transferred between memory and the processor [3,4]. Optical neural networks (ONNs) are expected to achieve orders of magnitude enhancement in energy efficiency and throughput due to their unique capability of all-optical matrix-multiplication at the speed of light without being subject to capacitive delays or loss due to heat dissipation [5–13]. However, three main challenges prevent ONNs from achieving competitive performance compared to digital ANNs. The first challenge is the lack of a device platform that can monolithically integrate optical neurons with lasers, photodetectors, electrical neural circuits, optical non-linear activation (NLA) functions, synaptic interconnects, and memory on a common silicon platform. The 2$^{nd}$ challenge is the lack of a clear solution to address power hungry optical weighting via thermal heaters which can consume 10s of mW each [14–16]. The 3$^{rd}$ challenge is the lack of optical memory which can potentially reduce communication latency between ONNs and memory. In light of this, there have been efforts to address integrated non-volatile photonics through the use of chalcogenide phase-change memory (PCM) [17–26], micro-electro-mechanical systems (MEMS) [27], ferroelectrics (BaTiO$_3$ [28], LiNbO$_3$ [29], PZT), floating-gate memory (FGM) [10,30–37], and memristors [38–42]. Memristors encounter several performance and manufacturability challenges which have prevented industry wide adoption; this includes performance variability, latency, density, and technological feasibility [43]. In the meantime, there is a need for a near term and reliable non-volatile silicon photonics solution such as charge-trap memory (CTM) or FGM [31]. Song et al., demonstrated FGM based non-volatile optical switching in 2016 via ring resonators. To the best of our knowledge, there has been only 1 demonstration of electrically driven CTM ring resonator albeit with large write voltage of V = 50 V and no electrical reset operation [36,44]. There have also been a few simulation studies outlining various configurations of CTM with photonics going back to 2006 [30] and more recently 2021 [32,35,45]. The difference between CTMs from FGMs is such that the charge trapping layer is an insulator instead of a conductor (poly-Si) [46–48]. There are 2 main disadvantages for using FGM: 1) stored electrons have a tendency to leak because of interfacial proximity to the tunnel oxide/floating gate region, and 2) high data write loads in FGMs can cause stress on the tunnel oxide layer, thus creating oxide defects which act as a leakage path from the poly-Si floating gate to the channel or source/drain regions [46,49,50]. CTM devices are immune to such failures since the floating gate consists of an insulator [46–48,50]. We address the aforementioned challenges by demonstrating a heterogeneous III-V/Si photonic platform capable of non-volatile optical functionality via the CTM effect. In addition, this platform is suitable for seamless integration of quantum dot (QD) comb lasers [51–54], III-V/Si MOSCAP ring modulators [55–57], Si-Ge avalanche photodetectors (APDs) [58–61], QD APDs [62,63], in-situ III-V/Si light monitors [64,65], III-V/Si MOSCAP optical filters [66,67], and non-volatile phase shifters [38–41,68], which are all essential towards realizing a fully integrated optical chip. We believe the co-integration of silicon photonics and non-volatile CTM memory provides a possible near term path towards eliminating the von-Neumann bottleneck as well as playing a role in energy efficient, non-volatile large scale integrated photonics such as: neuromorphic/brain inspired optical networks [5,6,15,26,69–74], optical switching fabrics for tele/data-communications [75,76], optical phase arrays [77,78], quantum networks, and future optical computing architectures. In this work, we demonstrate, for the first time, heterogeneous III-V/Si MZI and ring resonators with co-integrated CTM memory cells operating at O-band wavelengths. The MZI CTM exhibit full write/erase operation (100 cycles with 500 states) with wavelength shifts of $\Delta\lambda_{\text{non-volatile}}$ = 1.16 nm ($\Delta n_{\text{eff,non-volatile}}$ ~ 2.5 × 10$^{-4}$) and a dynamic power consumption < 20 pW (limited by measurement). The extinction ratio (ER) is ~ 1.78 dB, mainly limited by imperfect directional couplers. Multi-bit write operation (2 bits) is demonstrated and verified over a time duration of 24 hours and most likely beyond. The cascaded 2$^{\text{nd}}$ order ring resonator CTM filter exhibited an improved ER of ~ 7.11 dB compared to the MZI and wavelength shifts of $\Delta\lambda_{\text{non-volatile}}$ = 0.041 nm ($\Delta n_{\text{eff, non-volatile}}$ = 1.5 × 10$^{-4}$) with similar dynamic power consumption as the MZI CTM.

## Principle of Operation

CTM flash memory cells have typically been based on the SONOS (silicon-oxide-nitride-oxide-silicon) configuration where the tunneling, charge trap, and blocking layers are defined by SiO$_2$, Si$_3$N$_4$, and SiO$_2$ respectively [79]. High-k dielectric materials (HfO$_2$, ZrO$_2$, TiO$_2$) have become popular for the charge trap region because of increased potential barrier height ($\phi_0^{TiO_2}$ = 3.15, $\phi_0^{HfO_2}$ = 1.65 > $\phi_0^{Si_3N_4}$ = 1.03) which leads to improved charge retention times ( > 10 years) and improved programming speeds (~ μs) via effective oxide thickness (EOT) reduction [46,50]. Our optical CTM cell is based on an n-GaAs/Al$_2$O$_3$/HfO$_2$/Al$_2$O$_3$/HfO$_2$/Al$_2$O$_3$/Si heterogeneous III-V/Si structure, where the Al$_2$O$_3$ and HfO$_2$ serve as the tunneling/blocking oxide and charge trap respectively as shown in Fig. *1*a - b. In addition to high potential barriers, HfO$_2$ was chosen because of reported deep energy level traps (E$_t$ = 1.5 eV) [46,80] and high electron density traps ranging from 10$^{19}$ – 10$^{21}$ cm$^{-3}$ [32,49,81]. An Al$_2$O$_3$ layer is inserted in between the HfO$_2$, because it was experimentally determined to be easier to wafer-bond Al$_2$O$_3$ to Al$_2$O$_3$ rather than HfO$_2$. The choice of n-GaAs over p-GaAs was two-fold: 1) lower optical absorption loss from dopants, and 2) easier III-V/Si laser integration. Also, GaAs exhibits ~ 4 × smaller electron effective mass and ~ 6 × larger electron mobility (m$_e^*$ = 0.063m$_0$, μ$_e$ = 8500 cm$^2$/V-s) than crystalline Si (m$_e^*$ = 0.28m$_0$, μ$_e$ = 1400 cm$^2$/V-s) [63,67,82]. Therefore, the plasma dispersion effect on index change in n-type GaAs is more efficient with lower free carrier absorption (FCA) loss. The single-mode waveguide structure is defined by a width, height, and etch depth of 500, 300, and 170 nm respectively as indicated in *Fig. 3*a - b. The wafer-bonded III-V region is primarily 150 nm-thick n-GaAs doped at $3\times10^{18}$ cm$^{-3}$. *Fig. 3*c shows the simulated transverse electric (TE) of the optical CTM cell. Assuming dielectric thicknesses of 1.7/4.0/2.0/4.0/0.5 nm (*Fig. 1e – f*) and refractive indices of 1.75/1.90/1.75/1.90/1.75 for $Al_2O_3$/$HfO_2$/$Al_2O_3$/$HfO_2$/$Al_2O_3$ respectively, the calculated optical confinement factors are $\Gamma_{Si}$ = 64.49 %, $\Gamma_{HfO2}$ = 1.637 %, and $\Gamma_{Al2O3}$ = 0.82 % with an overall effective index of $n_{eff}$ = 3.0971 and group index of $n_g$ = 3.7914. For comparison, a pure silicon waveguide with oxide cladding has an effective index of $n_{eff}$ = 2.9774 and group index of $n_g$ = 3.9765.

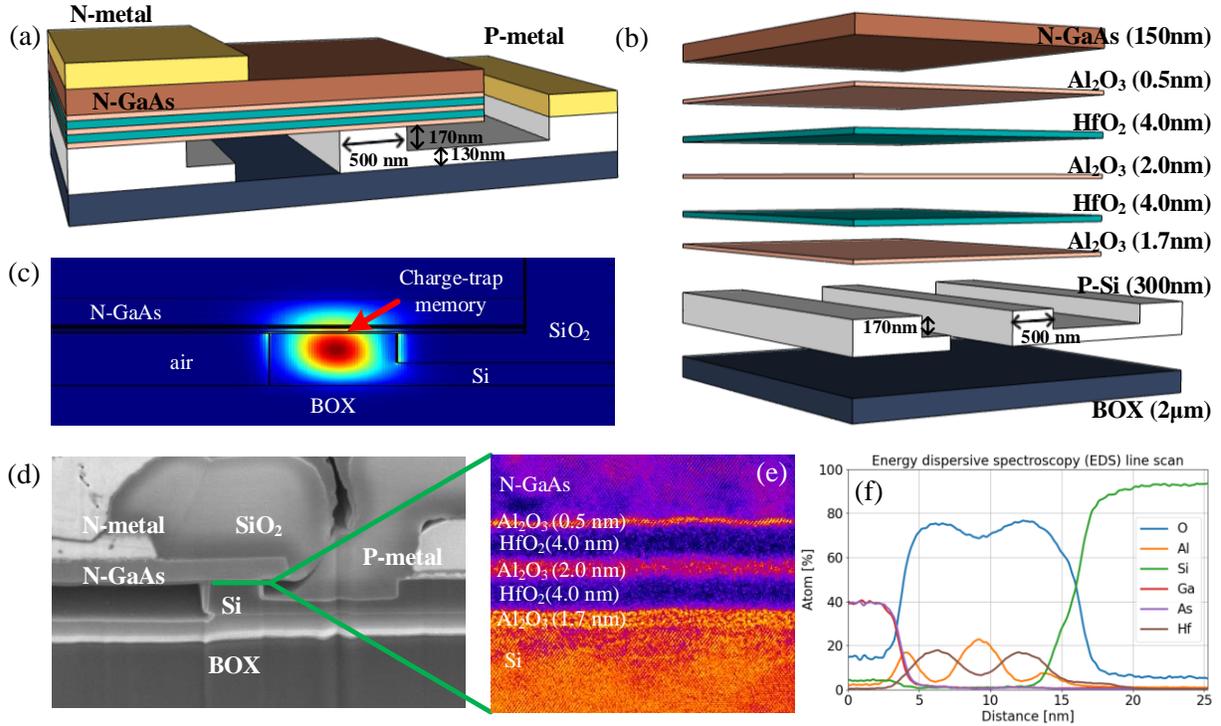

Fig. 1 (a) 3-D schematic of the heterogeneous III-V/Si CTM cell with (b) blown up material layer view and respective dimensions. (c) Simulated TE optical mode with CTM interface, (d) SEM cross section of CTM cell with (e) high resolution transmission electron microscope (HRTEM) image of charge-trap memory dielectric stack, and (f) electron dispersive spectroscopy (EDS) line scan for atomic mapping

In order to understand the non-volatile operation of the optical CTM cell, Fig. *2a – b* illustrate the energy-band diagrams of the flat-band and biased regimes respectively. The bandgap energy ($E_g$), electron affinity ($\chi$), valence/conduction band offsets (VBO/CBO), refractive index (n), trap state density ($N_{TC}$) and trap state energy ($\varphi_d$), are listed in supplementary note 1. During the write process (Fig. 2b), a positive bias is applied to the p-Si region which injects electrons from the highly doped n-GaAs into the high-k HfO$_2$ where carriers are trapped due to the presence of charge traps. These charge traps can exist in the bulk and interface regions, but for simplicity, we model the bulk trap case. Once the HfO$_2$ region is fully charged, holes will accumulate at the p-Si/Al$_2$O$_3$ interface, thus altering the effective index of the optical mode due to the plasma dispersion effect [83,84]. During the erase process, a reverse bias is applied to sweep out the trapped electrons, thus returning the optical CTM cell back to the initial electrical and optical state. A two-dimensional solver (SILVACO ATLAS [85]) was used to perform energy-band diagram and charge concentration calculations to theoretically predict optical effective index changes as a function of trapped charge density. The solver numerically calculates the Poisson and charge continuity equations and the effects of defect traps and self-consistently solves quantum mechanical tunneling. The main contributions to carrier injection are: Fowler-Nordheim tunneling, direct tunneling, and hot carrier injection.

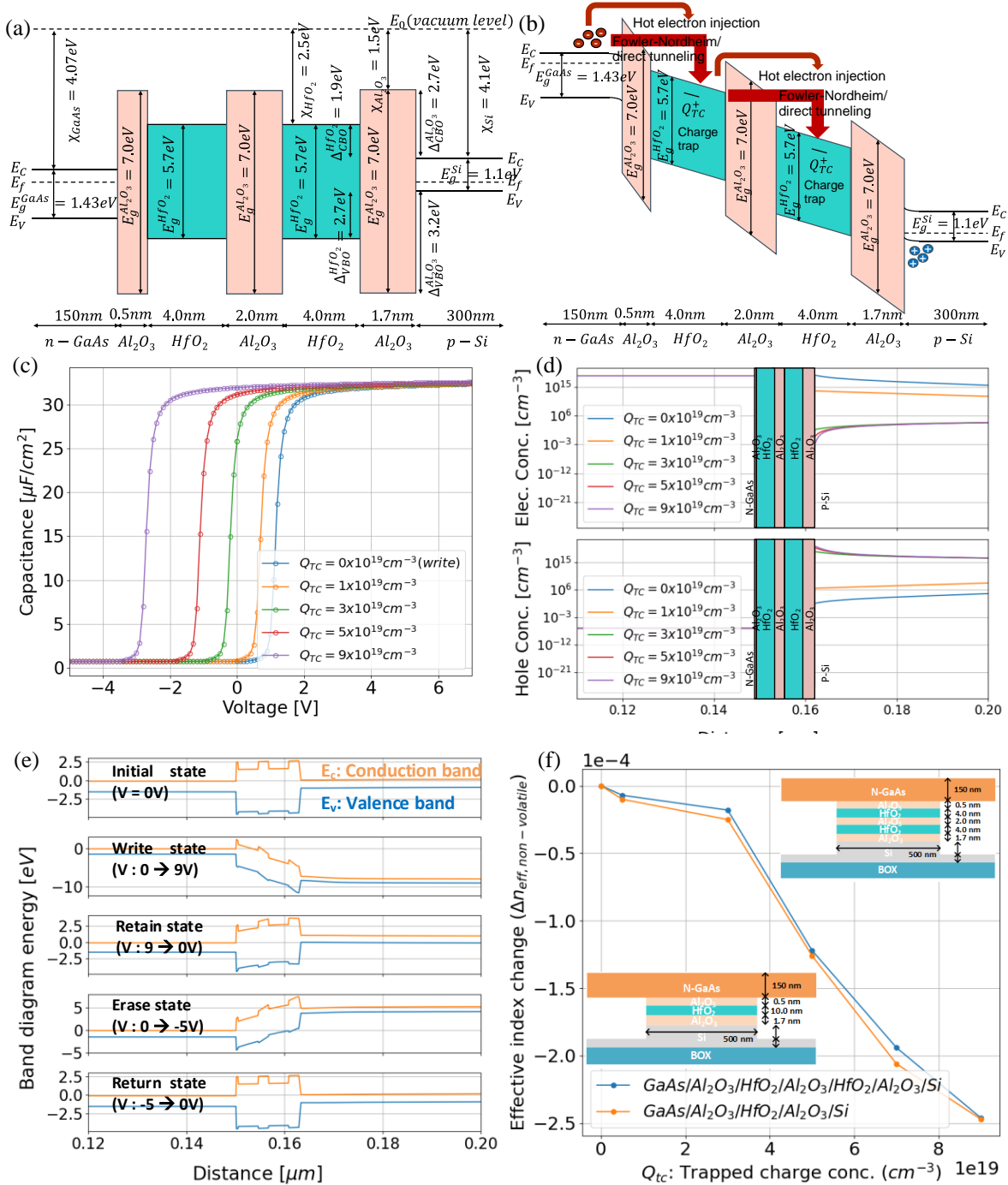

Fig. 2. Schematic of energy-band diagram for n-GaAs/$Al_2O_3$/$HfO_2$/$SiO_2$/p-Si CTM cell (a) at flat-band condition, and (b) with positive bias on the p-Si side. Simulated (c) C-V curves for various fixed charge concentration $Q_{TC}$, (d) electron and hole concentrations for various $Q_{TC}$ at retention state (0 → 9 → 0 V), (e) energy-band diagrams for 5 states: initial, write, retain, erase, and return, (f) non-volatile effective index change $\Delta n_{eff,non-volatile}$ vs. trapped charge concentration $Q_{TC}$ for different CTM dielectric stacks.

There have been reports of both acceptor and donor traps in $HfO_2$, however, the majority of research considers electron traps [46,86,87] and this is what we will consider in simulations. Non-volatile electrical behavior of CTM devices can be observed by the hysteresis of the capacitance-voltage (C-V) curve [46,88]. Fig. *2*c shows the simulated C-V hysteresis for our heterogeneous III-V/Si CTM structure with several values of trapped charge density ($Q_{TC}$) ranging from $0 - 10^{20}$ $cm^{-3}$. The upper limit of this range is consistent with reported $HfO_2$ density traps ranging from $10^{19} - 10^{21}$ $cm^{-3}$ [32,49,81]. The blue line represents the write state from the initial state (0 → 9 V), and the other colors represent a combination of retention (9 → 0 V), erase (0 → -5 V), and return (-5 → 0 V) states for different $Q_{TC}$. For example, if we assume a $Q_{TC} = 3 \times 10^{19}$, we would traverse the blue line from (0 → 9 V) for write operation, then the green line from (9 → 0 V) for retention, then green line from (0 → -5 V) for erase, and finally (-5 V → 0 V) for the return state. Next, the energy band diagrams for all 5 states are calculated (Fig. *2*e shows the case for $Q_{TC} = 3 \times 10^{19}$ $cm^{-3}$) along with the electron and hole concentrations in the retention state (Fig. *2*b). These values of electron and hole concentrations can be used to calculate a spatial change in index [83,84]: $\Delta n(x,y)$ (at 1310 nm) = $-6.2 \times 10^{-22} \Delta N(x,y) - 6 \times 10^{-18} \Delta P(x,y)^{0.8}$, where x and y are the 2D lateral and vertical dimensions as detailed in supplementary section. The resulting spatial indices are then used in an optical finite-difference-eigenmode (FDE) solver to calculate non-volatile effective index changes $\Delta n_{eff,non-volatile}$ vs. $Q_{TC}$ as shown in Fig. 2f. Two types of CTM structures are evaluated in Fig. *2*f; the structure used experimentally which entails 2 $HfO_2$ layers vs. 1 $HfO_2$ layer with an equivalent thickness. Because an equivalent thickness is used, it is expected that effective index changes will be similar, however, literature suggests temperature annealing (> 800 °C) is a solution for charge trapping enhancement at the $HfO_2/Al_2O_3$ interface [49,89]. The corresponding effective index changes on the order of $10^{-4}$ can significantly affect amplitude changes when implemented in resonant systems such as ring resonators. Consider the case of a heterogeneous III-V/Si CTM add-drop ring resonator with the following parameters: group index $n_{g, III-V/Si} = 3.7912$, radius R = 10.0 µm, power coupling coefficient $\kappa^2 = 0.01$, and internal loss of $\alpha_i = 3$ dB/cm. By using the appropriate ring resonator transfer function equations [90] and assuming a trapped charge density of $Q_{TC} = 9 \times 10^{19}$ $cm^{-3}$ ($|\Delta n_{eff,non-volatile}| = 2.47 \times 10^{-4}$) in the retain state, this change in index is enough to theoretically create a non-volatile wavelength shift of $\Delta \lambda_{non-volatile} = 0.085$ nm resulting in an ER > 14 dB. Alternatively, the hybrid III-V/Si CTM cell can be implemented in a traveling wave interference device such as a Mach-Zehnder interferometer (MZI). Consider the heterogeneous III-V/Si CTM MZI with the following parameters: effective index $n_{eff, III-V/Si} = 3.0971$, $n_{eff, Si} = 2.9774$, base length $L_{III-V/Si} = 325$ µm, optical path length

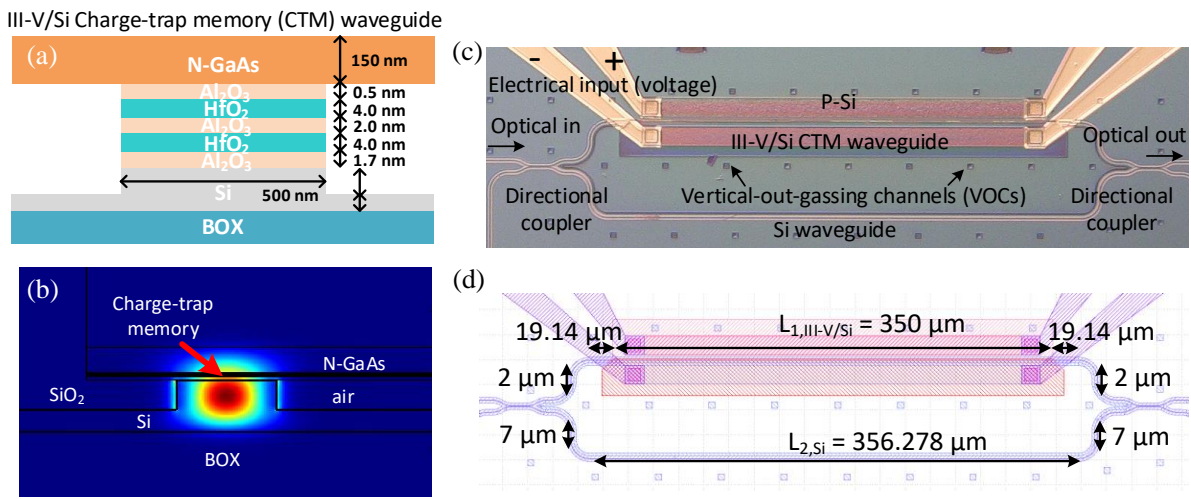

Fig. 3. (a) 2-D cross section of the heterogeneous III-V/Si CTM cell, (b) 2D-FDE simulated optical mode, (c) microscope image of fabricated III-V/Si CTM MZI, (d) design dimensions from mask layout

difference $\Delta OPL = 5.206$ μm, power coupling coefficient $\kappa^2 = 0.50$, and internal loss of $\alpha_i = 5$ dB/cm. These design parameters result in a free-spectral-range (FSR) of ~ 25.6 nm. Using the transfer matrix method for a MZI [91] and assuming the same trapped charge density of $Q_{TC} = 9 \times 10^{19}$ cm$^{-3}$ ($|\Delta n_{eff,non-volatile}| = 2.47 \times 10^{-4}$), this will result in a non-volatile phase shift of ~ $0.38\pi$ ($\Delta\lambda_{non-volatile} = 1.567$ nm) with an extinction ratio ER ~ 46 dB. The simulated electric field for the write process (V = 9 V) in the Al$_2$O$_3$/HfO$_2$/Al$_2$O$_3$/HfO$_2$/Al$_2$O$_3$ regions are $1.0 \times 10^7$, $3.75 \times 10^6$, $1.0 \times 10^7$, $3.75 \times 10^6$, $1.0 \times 10^7$ V/cm respectively. These are all well below experimentally reported breakdown electric field strengthens ($E_{BD, HfO2} = 4.80 \times 10^6$, $E_{BD, Al2O3} = 5 - 30 \times 10^6$ V/cm). Experimentally, we did not see breakdown when operating these devices as shown in the results section and supplementary section.

## Charge Trap Memory Experimental Demonstration

Initial phase tuning measurements were performed on a 350 μm long CTM Mach-Zehnder Interferometer (MZI) as shown in Fig. *3*d. Measured spectral response indicated an FSR ~ 16.58 nm with 1.62 nm of tuning at a 9V bias while maintaining an extinction ratio (ER) of ~ 10 dB. From transfer matrix modeling of the CTM MZI, the directional couplers were inferred to have a power transfer ratio of ~ 30 %, thus yielding the low ER. In order to investigate non-volatile CTM functionality, we applied a voltage cycle of [0, 9, 0, -5, 0] V and recorded the output optical spectra as shown in Fig. 4a. We start with the initial virgin state (blue curve) at 0 V and then proceed to bias up to 9 V which places the CTM MZI in a volatile state (orange curve) with a wavelength shift of $\Delta\lambda = 1.62$ nm. By turning off the voltage, a non-volatile state is reached as shown by the green curve with a non-volatile wavelength shift of $\Delta\lambda_{non-volatile} = 1.16$ nm. This translates to a non-volatile effective index change of $\Delta n_{eff, non-volatile}$ ~ $2.5 \times 10^{-4}$ which matches quite well with simulated results in Fig. 2f (for $Q_{TC} = 9 \times 10^{19}$ cm$^{-3}$). Simulations also show that perfect 50 % power couplers can yield an ER > 25 dB and 50 dB for waveguide losses of 53 and 3 dB/cm, respectively. A new fabrication run is currently underway with improved directional couplers and optical loss for the CTM MZIs. A reverse bias of -5 V perfectly resets this non-volatile state back to its original state (red curve).

We performed 100 voltage cycle tests (500 voltage states) to measure endurance/ repeatability performance and recorded optical spectra, resonant wavelength shifts, and I - V data, Fig. *4*c,e and Fig. *4*d,f, respectively. From Fig. *4*c, the mean wavelength resonances for the non-volatile and reset states are determined to be $1290.608 \pm 0.016$ nm and $1289.654 \pm 0.039$ nm respectively. This translates to write/erase wavelength accuracies of $\pm 2.807/6.842$ GHz which may be attributed to environmental temperature fluctuations. In parallel, the current-voltage relationship was tracked during optical measurements and is shown in Fig. *4*d. We believe measured current values are below the sensitivity of our measurement unit (Keithley 2400, 10 pA) which would indicate write/erase dynamic powers of < 20 pW. TEM images (supplementary note 4) of the initial (0 V), set (9 V), and reset state (- 5 V) indicate fully intact dielectric stacks with no visible signs of dielectric breakdown. EDS line scans also indicate minimal atomic/interfacial changes. It has been demonstrated that CTM cells exhibit charge retention times over 10 years [92–94].

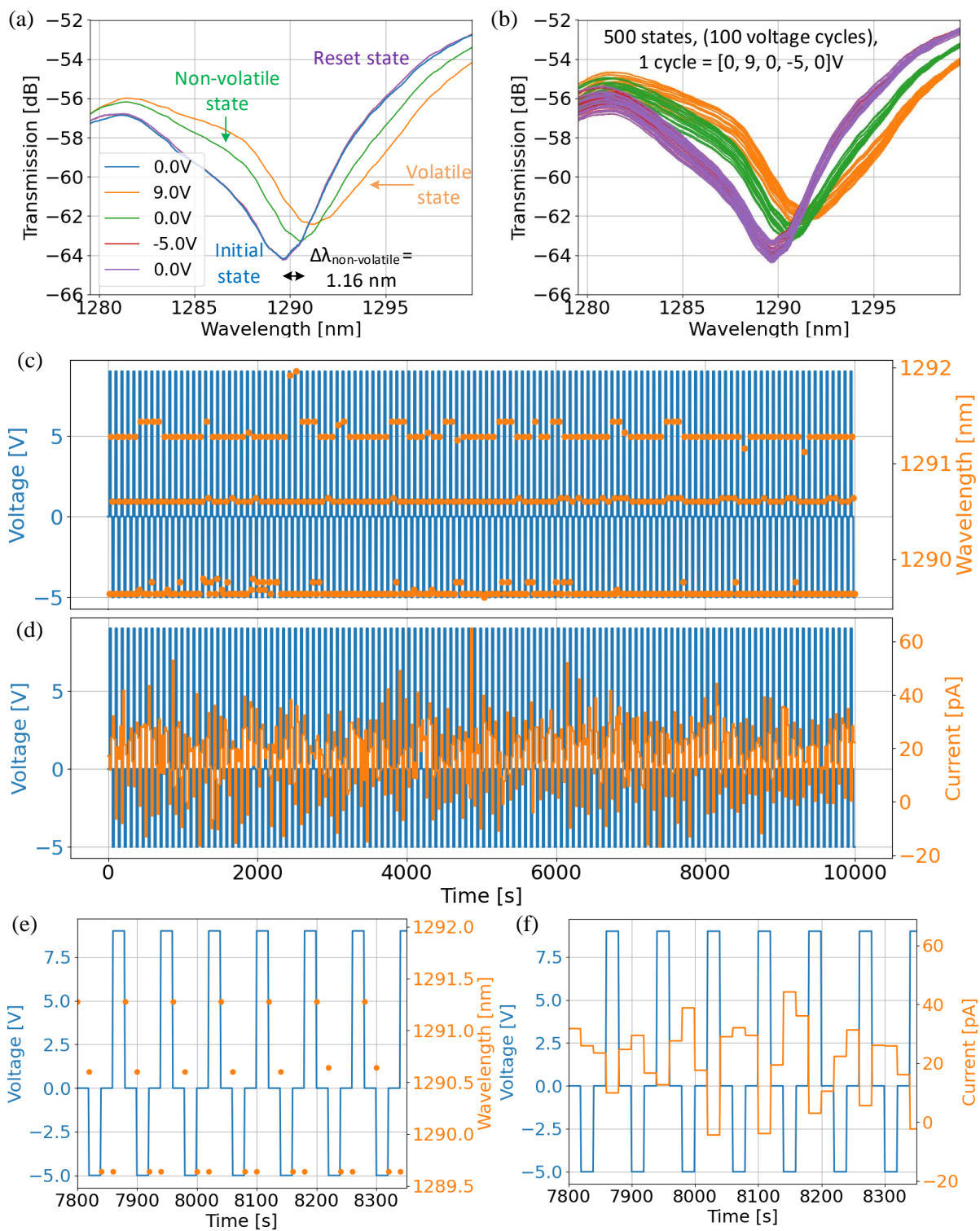

Fig. 4. Measured optical spectra for (a) initial state, volatile state, non-volatile state, reset state, final state, (b) 500 states via endurance testing: 100 voltage cycles with each cycle = [0, 9, 0, -5, 0]V. (c) Applied voltage vs. tracked resonance dip for 100 cycles. (d) Measured CTM voltage and current for 100 cycles. (e) – (f) Close-up snapshot of (c) – (d) respectively.

Preliminary study on the reliability of this CTM optical non-volatile state was conducted by performing a 24 hour time test for a particular "write" state as shown in Fig. 5a. The red curve is the minimum resonance wavelength for the initial optical response and the gray dots represent the resonant minima for the optical "write" state. The results indicate non-volatile states up to 24 hours and most likely beyond. Not shown here, but this particular "write" state held beyond 2 weeks before stopping experiments.

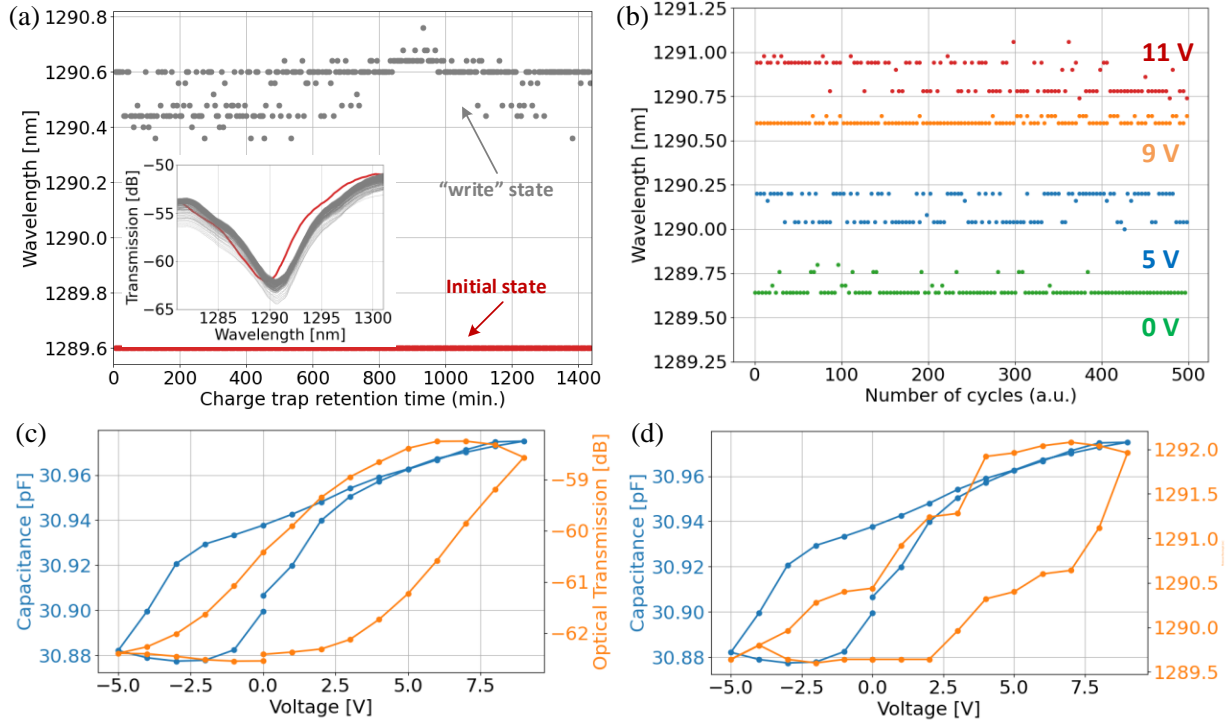

Fig. 5 (a) Measured optical "write" state tracked over a 24 hour period indicating non-volatile memory. (b) Demonstration of multi-bit optical states for 0, 5, 9, 11 V, (c) simultaneous C-V curve and non-volatile optical transmission difference at λ = 1289.57 nm, (d) simultaneous C-V curve and resonant wavelength minima.

Variance in the "write" wavelengths is observed and may possibly be attributed to room temperature variations due to the absence of a temperature controlled chuck. Recently, Zhang et al. have demonstrated electrical multi-bit (2 bits) information storage based on an $Al_2O_3/HfO_2/Al_2O_3$ gate stack with a record memory window exceeding 20 V [95]. In Fig. 5b, we demonstrate the optical analogue of multi-bit electrical storage for various applied voltage biases. We are able to achieve 2 bit retention with wavelength discretization errors due to the optical spectrum analyzer (OSA) resolution set to 0.2 nm. Taking into account the discretization errors, the mean and variance of write states 1 to 4 are $\bar{\lambda}_{\text{write}}$ = 1289.656, 1290.125, 1290.608, 1290.908 nm with a variance of $\sigma^2_{write}$= 1.6, 6.2, 0.25, 29.8 pm, respectively. The write states scale approximately linear for the first 3 states up until 11 V where there is an observed saturation in the wavelength red shift. Further improvement on bit resolution could be achieved by 4 solutions: 1) increasing Δn$_{\text{eff, non-volatile}}$ via engineering the number of trap defects to accommodate increased charge retention [46,79,86,96–98], 2) enabling larger optical mode overlap with the CTM cell, 3) making device longer, and 4) improved ER such that small wavelength shifts can still offer reasonable amplitude differences. Simultaneous C-V curves and optical transmission spectra were also performed to track the evolution of optical non-volatility. A hysteresis curve in Fig. 5c illustrates clear optical transmission non-volatility (at λ = 1289.57 nm) in the write to retention state (0 → +9 → 0 V). By applying an erase operation and observing the final state (0 → - 5 → 0 V), we can see near perfect reset of the device with a slight difference in amplitude of ~ 0.125 dB. Fig. 5d shows measured C-V curves while tracking resonant wavelength minima. Near perfect reset states are also achieved. The measured C-V curves indicate significant smearing compared to theoretical curves and can be attributed to interface states at the p-Si/SiO$_2$ and n-GaAs/HfO$_2$ interface. The interfacial trap density ($D_{it}$) can be extracted by employing the high-low frequency (*Castagné-Vapaille*) method [99,100] via the following equation: $D_{it} = (1/qA)[(C_{LF} - C_{ox})^{-1} - (C_{HF} - C_{ox})^{-1}]$, where $C_{LF}$ and $C_{HF}$ are the low and high frequencies measured at 20 kHz and 2 MHz respectively, A is the area, and q is unit charge. The calculated $D_{it}$ at near flat-band voltage is determined to be ~ $2.24 \times 10^{10}$ cm$^{-2}$ eV$^{-1}$, thus the existence of net positive charge trapped in the donor states [101] at the p-Si/SiO$_2$ can potentially reduce $\Delta n_{eff,non-volatile}$.

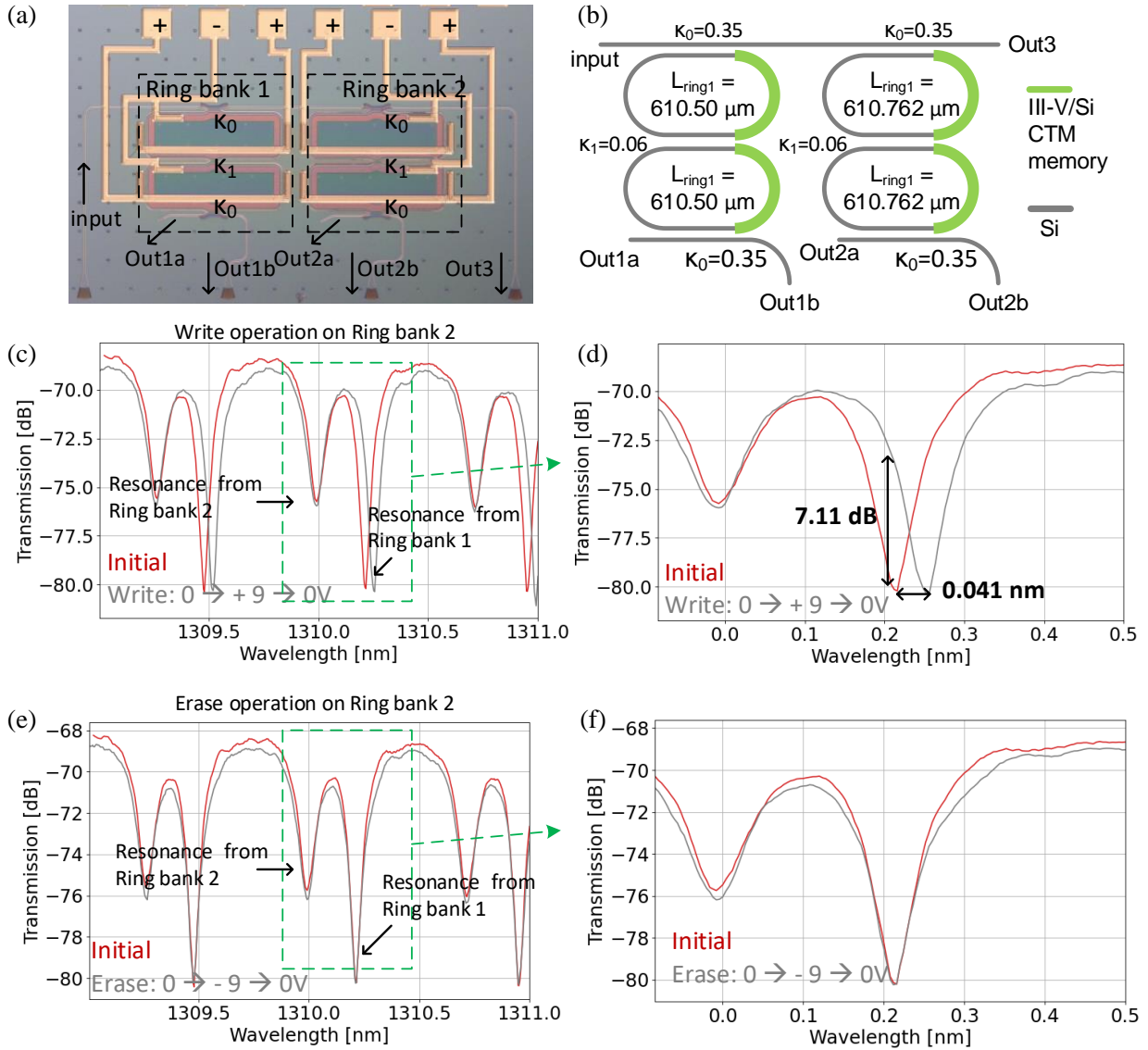

Fig. 6 (a) Microscope image of fabricated 2$^{nd}$ order cascaded ring filters with CTM cells, (b) design parameters, (c) Out3 optical spectrum for write operation (0 → + 9 → 0V) on ring bank 2, (d) close-up of spectrum indicating a 0.041 nm shift and a 7.11 dB extinction, (e) Output 3 optical spectrum for erase operation (0 → - 9 → 0V) on ring bank 2, (f) close-up of spectrum indicating a near perfect reset.

We have also integrated CTM cells into ring resonator-based designs which have improved ER over relatively small wavelength shifts. One example consists of two cascaded double ring structures as shown in Fig. 6a-b. The power coupling coefficients are chosen such that $\kappa_1^2 = 0.25k\kappa_0^4$ results in a maximally flat filter condition if k = 2 and $\kappa_0^2 = 0.35$ [102,103]. Each cascade double ring structure can be divided into "Ring bank 1" and "Ring bank 2", with the first ring bank designed for 130 GHz spacing assuming a group index of $n_g = 3.78$. Measurements yielded 126 GHz most likely due to fabrication imperfections and unknown experimental group indices. "Ring bank 2" is shifted by a $\Delta L = 26.25$ μm to offset resonances from "Ring bank 1" as shown in Fig. 6c-f. Fig. 6c-d shows the measured optical response before and after a non-volatile write operation of 0 → + 9 → 0 V. An ER of 7.11 dB was achieved with only a shift of $\Delta\lambda_{non-volatile} = 0.041$ nm. Transfer matrix calculations indicate an effective index change of $\Delta n_{eff} \sim 1.5 \times 10^{-4}$ nm (similar to MZI CTM result), which is quite comparable to plasma dispersion based phase shifters[83]. Duration tests were performed on this write state up to 24 hours yielding a mean wavelength shift of 0.027 nm with a variance of $6.7 \times 10^{-5}$ nm (supplementary note 3). Next, we proceeded to apply an erase voltage of 0 → - 9 → 0 V and Fig. 6e-f illustrates the near perfect reset in wavelength. Duration tests up to 24 hours were also performed for the erase state and yielded a mean wavelength shift of $- 4.30 \times 10^{-4}$ nm with a variance of $5.03 \times 10^{-5}$ nm (supplementary note 3). The current levels of both write and erase states were monitored to be fluctuating in the tens of pA indicating we are limited by the sensitivity of our measurement unit.

## Heterogeneous III-V/Si photonic in-memory computing platform

Our vision of co-located in-memory optical computing fabricated on a common heterogeneously integrated substrate is illustrated in Fig. 7. Fig. 7a shows a general artificial neural network (ANN) architecture composed of one input layer, N hidden layers, and an output layer. This ANN can be realized by our III-V/Si ONN where N layers can be achieved by time re-cycling the chip. This architecture is comprised of previously demonstrated HPE integrated photonic building blocks (III-V/Si QD lasers[104] and amplifiers[105], MOSCAP MZIs and microrings[67,106], lossless light monitors[64,65], QD APDs[62,107], SiGe APDs[58–60], programmable nonlinear activation functions[108,109], etc.). The non-volatile III-V/Si CTM cell in this work adds a final critical missing pieces that can be placed in MZIs or ring resonators without adding any design and fabrication complexity. The architecture in Fig. 7c uses MZIs, but can also be represented by a mesh of ring resonators [110]. The weights of an entire network can be first trained by using the III-V/Si MZIs in a *low-voltage, volatile* push-pull operation (up to ~ 30 Gbps in traveling wave electrode configuration [111]) while keeping track of the weight amplitudes with the III-V/Si lossless light monitor [64,65]. Fabrication imperfection and low-speed phase tuning due to environmental change can be compensated by the same MOSCAP as well, an athermal tuning process with negligible power consumption[63,67] unique to this platform. Fig. 7b shows a total tuning wavelength of 2.36 nm can be achieved from – 3 V to 2 V and with open 4 Gbps eye diagrams. In push-pull configuration, the wavelength tuning should be higher with equivalent drive voltage. The light monitors utilize internal trap mediated photo-carriers and induce no optical loss. These detectors can be part of a feedback circuitry with the non-volatile III-V/Si MZI CTM cell for true in-memory optical computing.

Once network training is complete, one can appropriately adjust one arm of the III-V/Si MZI using the higher voltage charge trap effect for non-volatile inference. Current optical non-volatile neuromorphic systems use PCM materials that require: 1) high power (4 mW) optical pulses (200 ns)[112] or 2) graphene thermal heaters in pulse operation (SET: 3 V, 100 μs, RESET: 5 V, 400 ns) to change from crystalline to amorphous and vice versa. Furthermore, these set/reset operations require a temporal separation in the seconds time range to ensure thermal relaxation[25]. Reconfiguring PCMs with optical pulses places additional scalability issues because number of light source and their controller equals to number of MZI in order to arbitrarily broadcast a network of control pulses into the system. In addition, ultrafast pulsed lasers can be quite power hungry and reduce MAC/J figure of merit. For the electrical based heating approach, the μs SET time and long enough thermal relaxation duration (2 s) between pulses can affect training throughput and suffer from thermal crosstalk. Our ability to train at current 10s of Gbps in low-power, non-volatile operation combined with reliable multi-bit non-volatile inferencing allows for increased throughput and energy efficiency which are lacking in PCM based approaches and modern day electrical MVM architectures.

The last but not the least, potentially integrated single- and multi-wavelength QD lasers offer convenience to architect the fan-out distribution through spatial or/and wavelength division multiplexing. The high-gain low-noise QD optical amplifiers boost up signals before or after reaching the inter-layer or final optical neurons. Thus, a fully-integrated, highly-scalable and programmable, energy-efficient photonic neuromorphic photonic chip including the entire optical computing functionality prior to inputting training outcome into a decision-making ASIC chip can be envisioned in the near future on this platform.

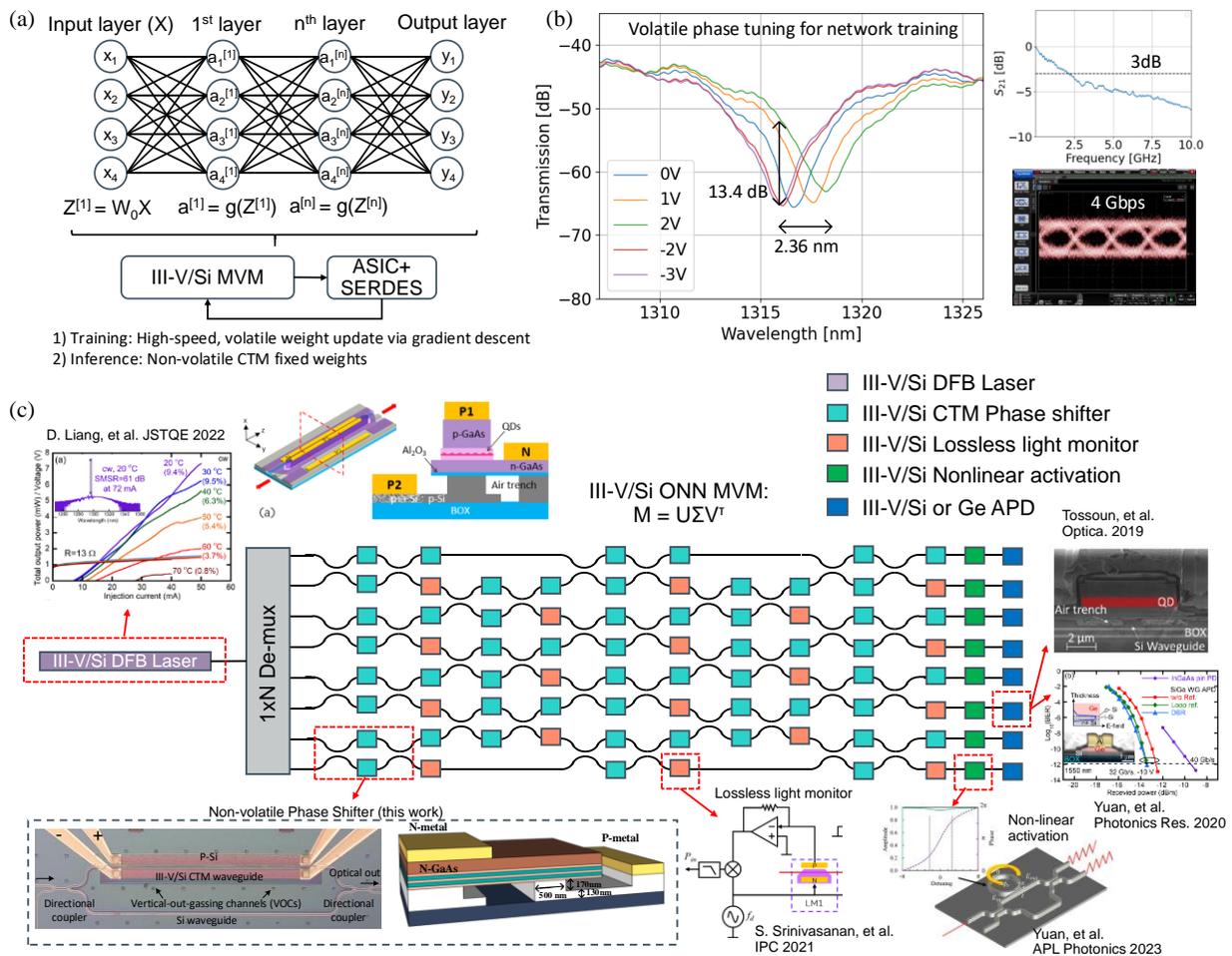

Fig. 7. (a) A general ANN architecture, (b) volatile phase tuning and 4Gbps speed measurements for network training, (c) schematic of a fully-integrated ONN MVM mesh on a heterogeneous III-V/Si platform.

## Conclusion

The von-Neumann bottleneck in conventional computing architecture inherently results in the need to transfer massive data between processor and memory with an intrinsic limit on bandwidth × distance plus increasing power consumption on the interconnect. As a major step towards breaking this bottleneck (especially for photonic neuromorphic computing), the work described here enables volatile operation for low-power, high-speed, on-chip training and non-volatile optical memory functionality for inference. This is all done on a heterogeneous III-V/Si platform capable of integrating all the necessary components needed for next generation applications such as: neuromorphic/brain inspired optical networks [5,6,15,26,69–74], optical switching fabrics for tele/data-communications [75,76], optical phase arrays [77,78], quantum networks, and future optical computing architectures. In particular, this work demonstrates for the first time, co-location of CTM memory cells with III-V/Si MZI and ring resonators which are key components in both optical communication and computing applications. The CTM memory cell is fully compatible with our heterogeneous III-V/Si platform and can offer benefits from an energy consumption, latency, and packaging standpoint. The MZI CTM exhibits full write/erase operation (100 cycles with 500 states) with wavelength shifts of $\Delta\lambda_{non-volatile}$ = 1.16 nm ($\Delta n_{eff,non-volatile} \sim 2.5 \times 10^{-4}$) and a dynamic power consumption < 20 pW (limited by measurement). The ER is ~ 1.78 dB, and can be improved to > 50 dB with the correct directional coupler design. Multi-bit write operation (2 bits) is also demonstrated and verified over a time duration of 24 hours and most likely beyond. The cascaded $2^{nd}$ order ring resonator CTM filter exhibited an improved ER of ~ 7.11 dB compared to the MZI and wavelength shifts of $\Delta\lambda_{non-volatile}$ = 0.041 nm ($\Delta n_{eff, non-volatile}$ = 1.5 × $10^{-4}$) with similar pW-level dynamic power consumption as the MZI CTM. We believe the demonstration and work presented in this paper can fuel further innovations in photonic co-integration with recent advanced CTM cells such as c-ZrTiO$_4$ (ZTO) [88], MoS$_2$/hBN/MoS$_2$/graphdiyne/WeS$_2$ [113], WSe$_2$/BN [114], etc.

## Data availability

The data that support the findings of this study are available from the corresponding author on reasonable request

## Competing interests

The authors declare no competing interest.

## Acknowledgements

We thank funding from DOE ARPA-E ULTRALIT contract No. DE-AR0001039, and USG MPO contract No. H98230-18-3-0001. We thank the UCSB nanofabrication facilities. We also thank Sung-Won Kong and Garrett Schlenvogt from Silvaco, Inc. for providing guidance and support with the ATLAS simulation solver.

## Author contributions

S.C. conceived the initial concept, and designed devices. D.L. designed the MOSCAP structure, fabrication flow, and was heavily involved in data analysis and computing architecture design. G.K. and Y.H. fabricated the devices and suggested improvements in the design phase. S.C., Y.Y., and Y.P. conducted the chip testing. D.L. and R.B. managed the project and gave important technical advice. All authors reviewed the manuscript.

# Supplementary

Stanley Cheung[*], Di Liang[*], Yuan Yuan[*], Yiwei Peng, Yingtao Hu, Geza Kurczveil, and Raymond G. Beausoleil

Hewlett Packard Enterprise, Large-Scale Integrated Photonics Lab, Milpitas, CA. 95035, USA
[*]stanley.cheung@hpe.com, di.liang@ieee.org, yuan.yuan@hpe.com

## Supplementary Note 1:

### Electro-optical simulations and design

A 2-D electrical solver (SILVACO ATLAS [7]) was used to study the non-volatile electrical behavior of the III-V/Si CTM structure. This modeling software has been used successfully in other theoretical optical non-volatile structures[8,9]. The program numerically solves the Poisson, charge continuity equations, drift-diffusion transport, and quantum tunneling mechanisms. The models involved include Fermi-Dirac statistics, Shockley-Read-Hall recombination, quantum tunneling that includes direct and Fowler-Nordheim [7]. The semiconductor material parameters used for optical and electronic TCAD simulations are listed in Table *1*: Material parameters used in electro-optical simulations.

Table 1: Material parameters used in electro-optical simulations

| Material | $E_g$ (eV) | $\chi$ (eV) | VBO (eV) | CBO (eV) | n | $N_{TC}$ (cm$^{-3}$) | $\varphi_d$ (eV) |
|---|---|---|---|---|---|---|---|
| $Al_2O_3$ | 7.0 [10] | 1.5 [11–13] | 3.2 [10] | 2.7 [10] | 1.75 | - | - |
| $HfO_2$ | 5.7 [10] | 2.5 [10] | 2.7 [10] | 1.9 [10] | 1.9 | $10^{19}$ - $10^{20}$ [9,14,15] | 2.0 [16–19] |
| $SiO_2$ | 8.9 [10] | 1.3 [10] | 4.5 [10] | 3.3 [10] | 1.44 | - | - |
| Si | 1.1 [10] | 4.1 [10] | - | - | 3.507 [20] | - | - |
| GaAs | 1.43 [7] | 4.07 [7] | - | - | 3.406 [20] | - | - |

$E_g$: Energy gap, $\chi$: electron affinity, n: refractive index (at $\lambda$ = 1310 nm), VBO: valance band offset, CBO: conduction band offset, $N_{TC}$: trap density, $\varphi_d$: electron trap level

Once the structure is built with the appropriately assigned models and trapped charge density, the band diagrams, C-V curves, and electron/hole concentrations are simulated for various biases. The spatial profiles of the change in electron/hole concentrations ($\Delta P(x,y)$ and $\Delta N(x,y)$) are then exported and used to calculate a spatial refractive index change $\Delta n(x,y)$ by the following equation [21,22]:

$$\Delta n(x, y) = -6.2 \times 10^{-22} \Delta N(x, y) - 6.0 \times 10^{-18} \Delta P(x, y) \quad (1)$$

This equation is valid for a wavelength $\lambda$ = 1310 nm and is used throughout the manuscript. Next, this spatial refractive index change $\Delta n(x,y)$ is exported into a 2D-FDE optical mode solver (Lumerical) and the non-volatile change in effective index ($\Delta_{neff,non-volatile}$) is calculated. This change can then be used to calculate non-volatile phase shifts on photonic devices such as ring resonators, Mach-Zehnder interferometers, filters, etc. The figure below illustrates the simulation procedure:

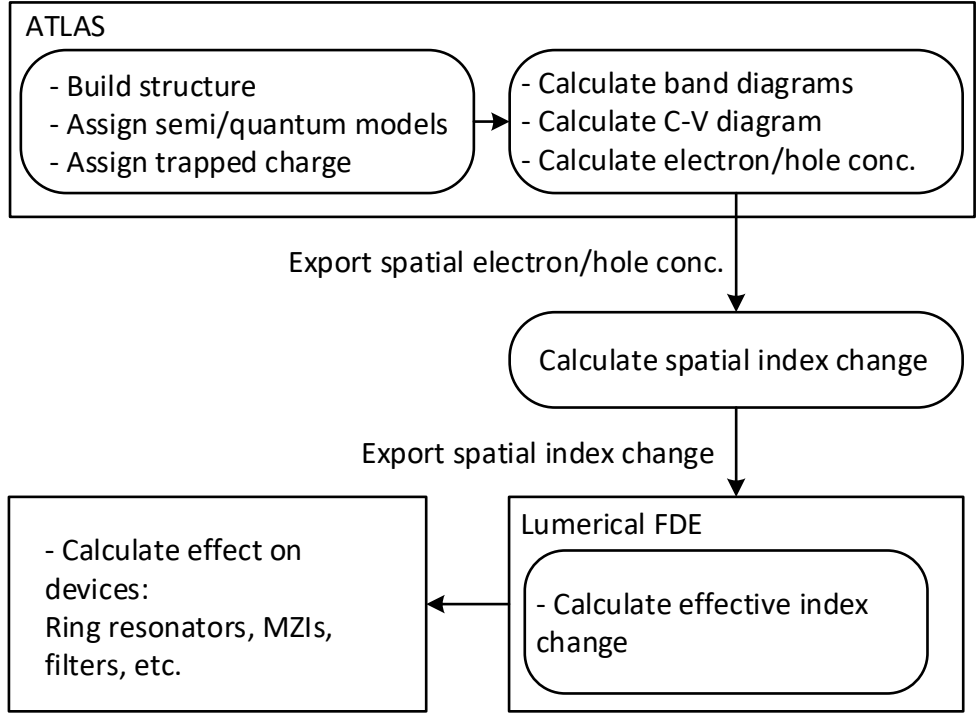

Fig. 8. Flowchart of electro-optical simulation procedure for III-V/Si CTM structure

Post-fabrication phase error correction

An alternative use of the non-volatile III-V/Si CTM cell is the role it can play in post-fabrication trimming or permanent phase error correction without consuming static electrical power. Inherent to the high index contrast of the silicon photonic material system, phase sensitive devices such as arrayed waveguide gratings (AWGs), lattice filters, and (de-)interleavers are sensitive to phase errors and are dependent on waveguide width, thickness, and refractive index non-homogeneity. In this case, the ability to use non-volatile phase tuning to correct the aforementioned errors is essential to minimizing power consumption. Changes in resonant the resonant wavelength can be described by the following equation [4]:

$$\Delta\lambda_0 = (\lambda_0 / n_g)\sqrt{(dn_{eff}/dw \cdot \Delta w)^2 + (dn_{eff}/dt \cdot \Delta t)^2} \tag{12}$$

$\lambda_0, n_{eff}, n_g, \Delta w,$ and $\Delta t$ are the free-space wavelength, effective index, group index, width variation, and thickness variation respectively. Along with the group index $n_g = n_{eff} - \lambda_0 \cdot dn_{eff}/d\lambda$, the resonant wavelength variation for each dimension can be calculated as: $\Delta\lambda_0 / \Delta w = (\lambda_0 / n_g)(dn_{eff}/dw)$ and $\Delta\lambda_0 / \Delta t = (\lambda_0 / n_g)(dn_{eff}/dt)$. The effective index and group index as a function of width and thickness are plotted in Fig. 9a – b. It can be seen that both values increase as waveguide dimensions increase because of increased modal confinement. Fig. 9c – d illustrates width sensitivity ($dn_{eff}/dw$) and thickness sensitivity ($dn_{eff}/dt$). Throughout the paper, single-mode III-V/Si CTM waveguides are used and have design dimensions of height = 300 nm, width = 500nm, etch depth = 170nm, and GaAs thickness of 150 nm, thus resulting in effective index variations of $dn_{eff}/dw = 5.80 \times 10^{-4}$ /nm and $dn_{eff}/dt = -4.44 \times 10^{-5}$ /nm. Typically, the most critical parameter in controlling phase errors is the starting SOI wafer thickness uniformity.

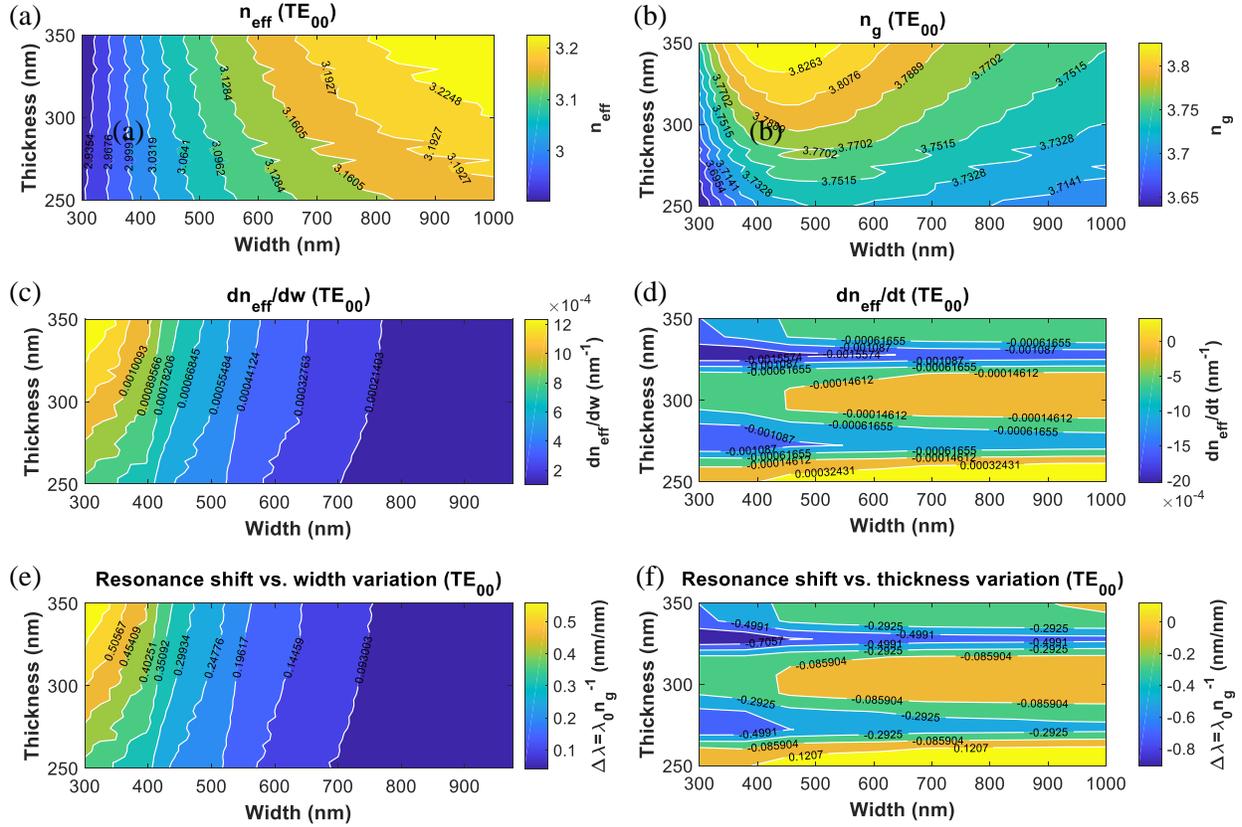

Fig. 9 III-V/Si CTM optical mode calculations for (a) effective index ($n_{eff}$), (b) group index ($n_g$), (c) effective index change vs. waveguide width ($dn_{eff}/dw$), (d) effective index change vs. waveguide thickness ($dn_{eff}/dt$), (e) wavelength shift ($\Delta\lambda_0/dw$) vs. waveguide width, (f) wavelength shift vs. waveguide thickness ($\Delta\lambda_0/dt$).

However, for the III-V/Si case, the GaAs thickness reduces this SOI thickness sensitivity significantly, and as a result, width variations play a larger role in phase errors by an order of magnitude. The wavelength shift variation are $\Delta\lambda_0/dw$ = 0.2457 nm/nm and $\Delta\lambda_0/dt$ = -0.0187 nm/nm as shown in Fig. 9e – f. The use of wider waveguides can significantly reduce $\Delta\lambda_0/dw$ by another order of magnitude, however, the TE01 mode starts to appear at a width of 600 nm as seen in Fig. 9b. We believe, the III-V/Si CTM waveguide dimensions used throughout this paper offers the best design trade-off in terms of low $dn_{eff}/dw$ and $dn_{eff}/dt$ while maintaining single-mode operation. However, if we assume a charge trap concentration of $Q_{TC} > 9 \times 10^{19}$ cm$^{-3}$ and a $dn_{eff}/dw = 5.80 \times 10^{-4}$ /nm, the III-V/Si CTM cell may be capable of correcting several nm of errors in waveguide width.

## Supplementary Note 2: Fabrication

In-house device fabrications starts with a 100 mm SOI wafer which consists of a 350 nm thick top silicon layer and a 2 μm buried oxide (BOX) layer. The top silicon is thinned down to 300 nm by thermal oxidation and buffered hydrofluoric (HF) acid etching, thus leaving a clean silicon surface. Silicon waveguides are defined by a deep-UV (248 nm) lithography stepper and boron is implanted to create p ++ silicon contacts. Grating couplers, silicon rib waveguides, and vertical out-gassing channels (VOCs) are respectively patterned using the same deep-UV stepper and then subsequently etched 170 nm with Cl$_2$-based gas chemistry. Next, the silicon wafer goes through a Piranha clean followed by buffered hydrofluoric (HF) acid etching to remove any hard masks. Next, an oxygen plasma clean is performed followed by a SC1 and SC2 clean and HF dip. The III-V wafer goes through a solvent clean consisting of acetone, methanol, and IPA, followed by oxygen plasma cleaning and a short dip in $NH_4OH:H_2O$ solution. Next a dielectric of $Al_2O_3$ is deposited onto both GaAs and Si wafers via atomic layer deposition (ALD) with a target thickness of 0.5 nm on each side followed by 3 nm $HfO_2$ and 1 nm $Al_2O_3$ depositions. The two samples are then mated and annealed at 250 °C . After thermal anneal, the III-V substrate is selectively removed by a combination of mechanical lapping and selective wet etching in $H_2O_2:NH_4OH$ solution.. Upon removing AlGaAs wet stop layer in buffered hydrofluoric acid (BHF) solution, n-GaAs surface is exposed . Ge-based n-contact metallization process is conducted beforethe III-V mesa is defined and dry etched to expose Si surface and metallization process on p++ Si. Next, a plasma enhanced chemical vapor deposition (PECVD) $SiO_2$ cladding is deposited and the vias are defined and etched. Finally, thick metal probe pads are defined to make contact with n- and p-contacts. The relevant fabricated devices can be seen in Fig. *11*a-b.

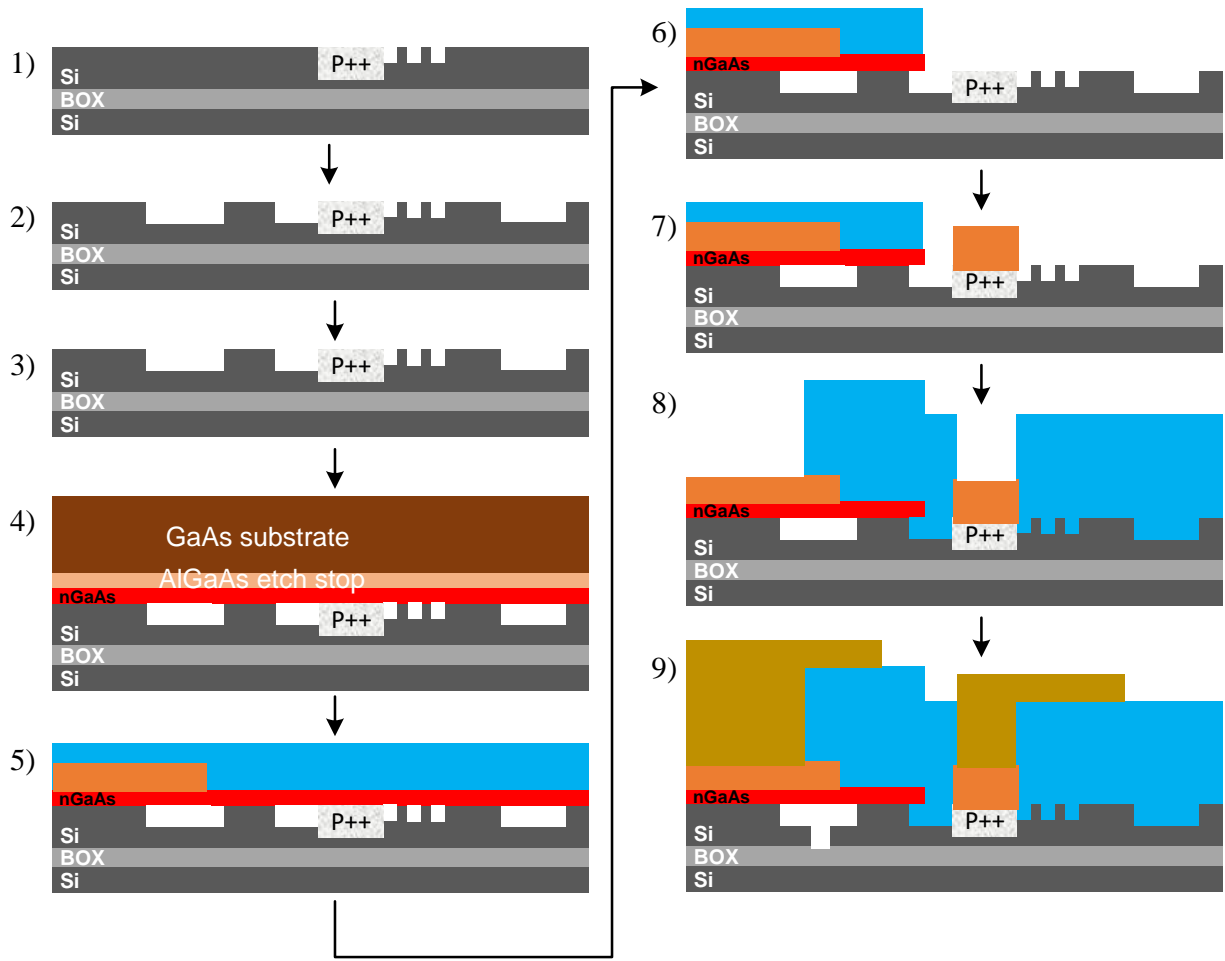

Fig. 10 Fabrication flow of heterogeneous III-V/Si CTM device.

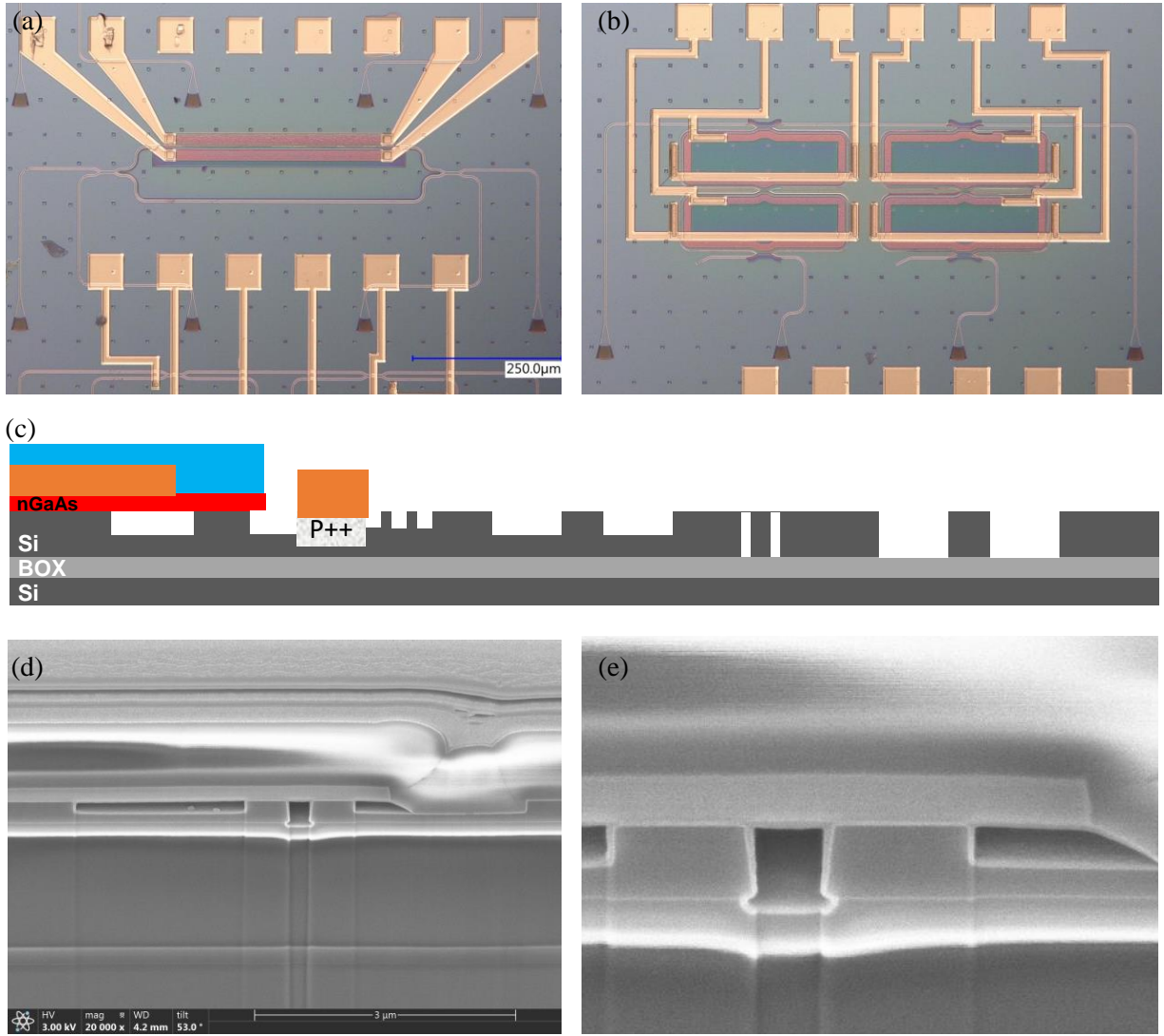

Fig. 11. Microscope images of fabricated III-V/Si CTM (a) MZI, and (b) 2$^{nd}$ order cascaded ring resonators. (c) Schematic of 2-D cross-section. (d) - (e) SEM image of cross-section.

## Supplementary Note 3: Measurements

The spectral response of the devices were characterized with a Thorlabs superluminescent diode (SLD) capable of 40 nm bandwidth (1290 – 1330 nm). The 100 mm wafer is vacuum mounted onto a stainless steel chuck on a semi-automatic probe station. Light is vertically coupled in/out of devices via grating couplers with a 7 ° polished fiber array. Polarization control is performed with the use of a polarization controller and maximized for peak transmission on a straight test waveguide. C-V, I-V, and optical spectra measurements are performed with an Agilent E4980A, Keithley 2400, and Yokogawa AQ6370D respectively. The pre-bonded 0.5 μm wide straight Si waveguide TE losses were determined to be about ~ 9.2 dB/cm primarily due to sidewall roughness for a wavelength of 1310 nm from a series of cutback test structures. 0.8 μm wide multi-mode straight Si waveguide TE losses were about 9.8 dB/cm. Circular bends of radius = 2, 5, 7, 14 μm had bend losses = 1.22, 0.83, 0.3, 0.08 dB/90$^o$ bend respectively. Grating coupler loss before and after bonding were calculated to be 7.7 and 7.8 dB/coupler indicating negligible effect after III-V removal.

*Fig. 12*a shows the measured transmission spectra of the 2nd order cascaded ring resonator filter. The filter shape is far from ideal and transfer matrix modeling indicates power coupling coefficients of $\kappa_0 = 0.43$, $\kappa_1 = 0.02$ for "Ring bank 1" and $\kappa_0 = 0.43$, $\kappa_1 = 0.09$ for "Ring bank 2" as shown in Fig. *12*b. By applying a write operation (0 → + 9 → 0 V), a $\Delta\lambda_{non-volatile} = 0.041$ is achieved which corresponds to a $\Delta n_{eff,\ non-volatile} = 1.5 \times 10^{-4}$. This change in effective index is similar to the MZI CTM result ($\Delta n_{eff,\ non-volatile} = 2.5 \times 10^{-4}$).

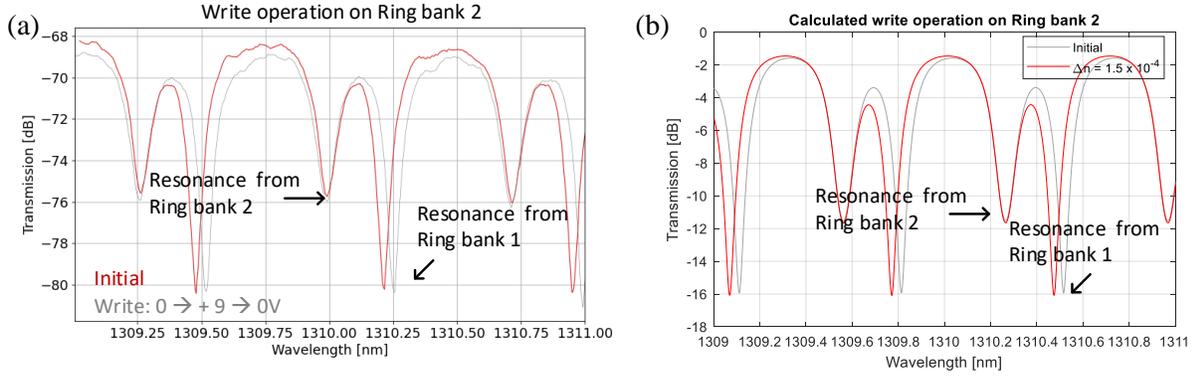

Fig. 12 III-V/Si CTM 2nd order cascaded ring resonator filters: (a) measured, and (b) calculated. Measured $\Delta\lambda_{non-volatile} = 0.041$ nm corresponds to a $\Delta n_{eff,\ non-volatile} = 1.5 \times 10^{-4}$.

Non-volatile optical retention measurements for the "write" and "reset" states were performed over a 24 hour period and are shown in Fig. *13*a, and b respectively.

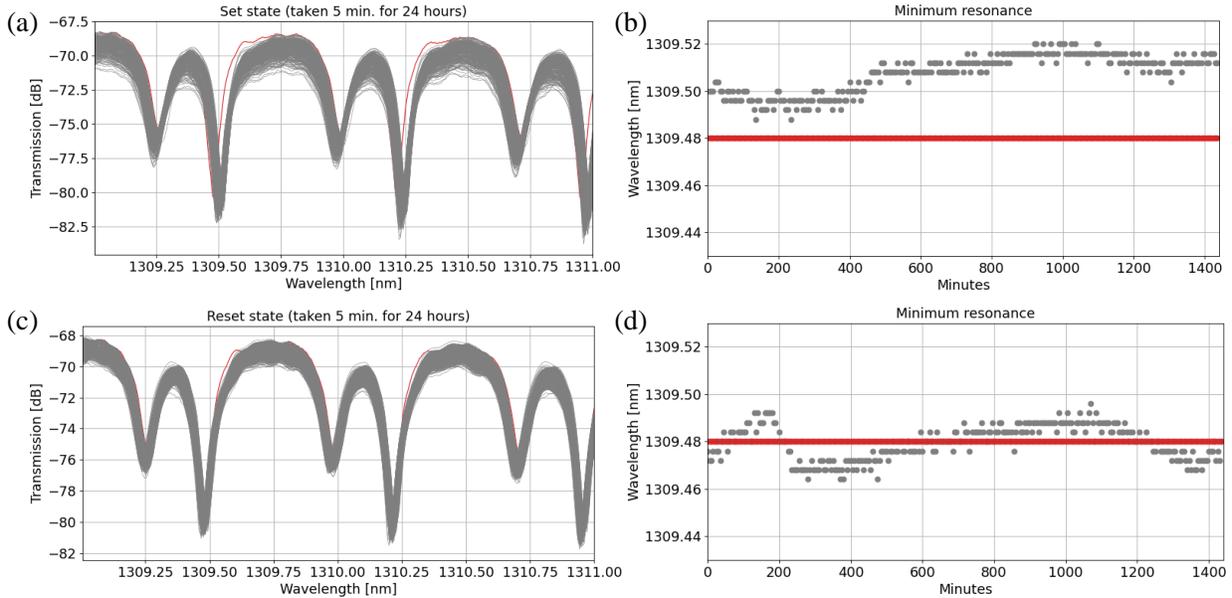

Fig. 13 24 hour data tracking for (a) "write" state optical spectra, (b) non-volatile wavelength shift for a resonant minimum. (c) "Reset" state optical spectra, and (d) non-volatile wavelength reset.

## Supplementary Note 4: Transmission electron microscope (TEM) measurements of initial, set, and reset states.

We performed transmission electron microscope (TEM) imaging of initial, set, and reset states as shown in *Fig. 14*a-b respectively. Each image is a separate sample that was biased and subsequently imaged. There does not appear to be any visual evidence of dielectric breakdown in the n-GaAs/Al$_2$O$_3$/HfO$_2$/Al$_2$O$_3$/HfO$_2$/Al$_2$O$_3$/Si CTM structure. There are contrast differences and may indicate variability in sample preparation as well as imaging settings. The electron dispersive spectroscopy (EDS) line scans below the TEM images also indicate minimal atomic/interfacial changes.

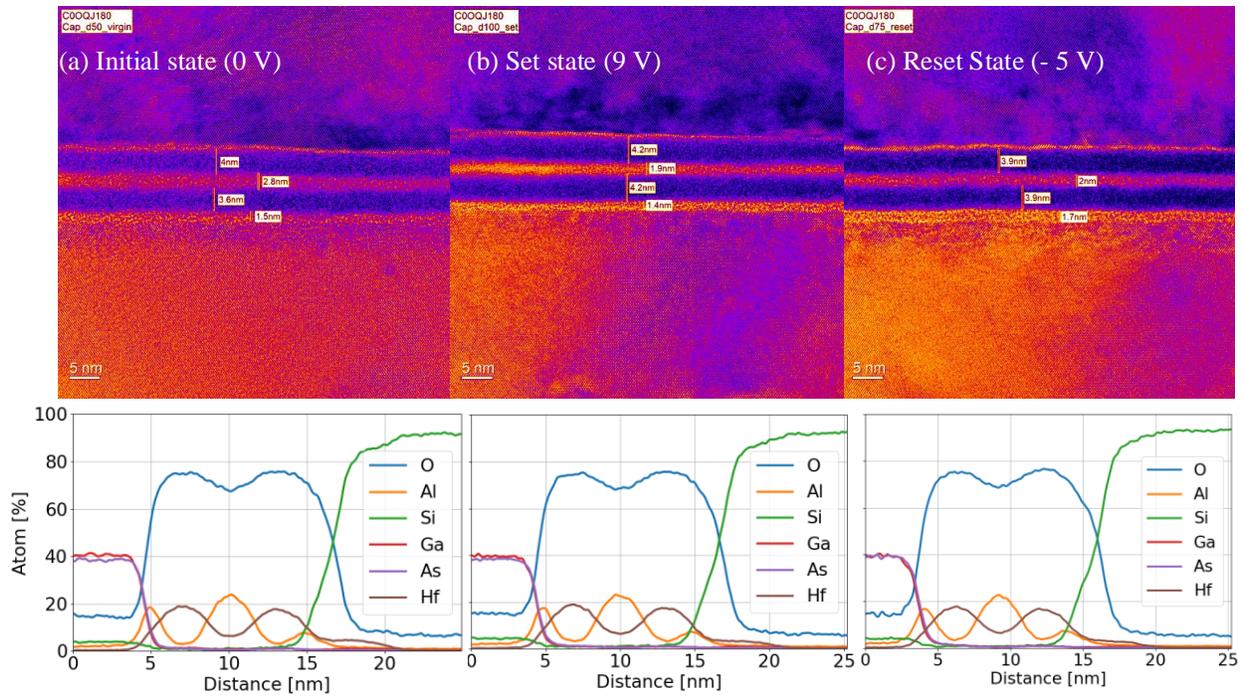

Fig. 14 TEM imaging and EDS line scan of n-GaAs/Al$_2$O$_3$/HfO$_2$/Al$_2$O$_3$/HfO$_2$/Al$_2$O$_3$/Si CTM structure in (a) initial state (0 V), (b) set state (9 V), and (c) reset state (- 5 V).